\documentclass[%
 reprint,
 amsmath,amssymb,twocolumn,prl
]{revtex4-2}

\usepackage{graphicx}
\usepackage{dcolumn}
\usepackage{bm}
\usepackage{makecell}
\usepackage[sort&compress]{natbib}
\begin{document}

\preprint{APS/123-QED}

\title{Transverse Bending Mimicry of Longitudinal Piezoelectricity}

\author{Zhi Tan}
 \email{tanzhi0838@scu.edu.cn}
 \affiliation{College of materials science and engineering, Sichuan University, Chengdu 610065, China.}

\author{Xiang Lv}
\affiliation{College of materials science and engineering, Sichuan University, Chengdu 610065, China.}
 
\author{Jie Xing}%
 \email{xingjie@scu.edu.cn}
  \affiliation{College of materials science and engineering, Sichuan University, Chengdu 610065, China.}
  
\author{Shaoxiong Xie}
\affiliation{Department of Materials Science and Engineering, Friedrich-Alexander-Universität Erlangen-Nürnberg (FAU), Erlangen 91058, Germany}

\author{Hui Zhang}
\email{zhanghui@xidian.edu.cn}
\affiliation{
 Shaanxi Key Laboratory of High-Orbits-Electron Materials and Protection Technology for Aerospace, School of Advanced Materials and Nanotechnology, Xidian University; Xi'an 710126, China 
}%

 \author{Jianguo Zhu}%
 \email{nic0400@scu.edu.cn}
\affiliation{%
 College of materials science and engineering, Sichuan University, Chengdu 610065, China\\
}%

\date{\today}

\begin{abstract}
The origin of frequently observed ultrahigh electric-induced longitudinal strain, ranging from 1\% to 26\%, remains an open question. Recent evidence suggests that this phenomenon is linked to the bending deformation of samples, but the mechanisms driving this bending and the strong dependence of nominal strain on sample thickness have yet to be fully understood. Here, we demonstrate that the bending in piezoceramics can be induced by non-zero gradient of \textit{d}\textsubscript{31} acrcoss thickness direction. Our calculations show that in standard perovskite piezoceramics, such as KNbO\textsubscript{3}, a 0.69\% concentration of oxygen vacancies results in a 6.3 pC/N change in \textit{d}\textsubscript{31} by inhibiting polarization rotation, which is sufficient to produce ultrahigh nominal strain in thin samples. The gradients of defect concentration, composition, and stress can all cause sufficient inhomogeneity in the distribution of \textit{d}\textsubscript{31}, leading to the bending effect. We propose several approaches to distinguish true electric-induced strain from bending-induced effects. Our work provides clarity on the origin of nominal ultrahigh electric-induced strain and offers valuable insights for advancing piezoelectric materials. 

\end{abstract}

\maketitle


The electromechanical coupling, i.e. piezoelectricity, enables inter-conversion of mechanical and electric information and energy\cite{uchino1996piezoelectric,uchino2015glory}. This unique property underpins a wide range of applications for piezoelectric materials, including acceleration sensors\cite{accelerometers}, actuators\cite{PiezoelectricActutors,PiezoelectricActuators-Gao}, ultrasonic medical imaging\cite{MedicalImaging}, and emerging technologies like piezoelectric energy harvesting\cite{PiezoEnergyHarvesting}. For actuators, achieving high strain under an applied electric field is crucial for enhancing their sensitivity and performance. Despite various strategies, such as constructing morphotropic phase boundaries\cite{PiezoelectricCeramics}, creating nanoscale structural heterogeneity\cite{LiFei-AFM-heterogeneity,li2019-science}, and etc., are developed, the conventional piezoelectric strain usually only reach 0.1\% $\sim$ 0.3\% in perovskite ferroelectrics\cite{li2019-science,saito2004lead}. Recent advancements have led to modified Bi\textsubscript{0.5}Na\textsubscript{0.5}TiO\textsubscript{3}\cite{zhang2007giant} and BiFeO\textsubscript{3}-BaTiO\textsubscript{3}\cite{DawangWang-PRL} based ceramics that can exhibit high strains exceeding 0.4\%, achieved through a phase transition from an ergodic relaxor state to a ferroelectric state under an applied electric field. However, to trigger this type of strain, usually a very high external electric field is required and accompanied by large hysteresis\cite{zhang2007giant}, which are not favored in practical applications. 

Recently, remarkable high strains (1\% $\sim$ 26\%) have been observed in several films\cite{park2022-CeO2,lin2024ultrahigh-NN,pan2024clamping, waqar2022origin-NC} and piezoceramics\cite{KNN2022gian-Scinece, luo2023achieving-SA, wang2023giant-AFM}, mostly containing artificially introduced defects. Notably, most of high strains are typically achieved in samples with thin thicknesses, often below 0.3 mm. Interestingly, it has been reported that this kind of electric-induced strain is strongly dependent on the thickness of piezoceramic disks\cite{adhikary2023ultrahigh-JAP}. For example, at a thickness of 0.22 mm, K\textsubscript{0.49}Na\textsubscript{0.49}NbO\textsubscript{3} piezoceramic with defects exhibits a nominal longitudinal strain of approximately 3.5\%\cite{adhikary2023ultrahigh-JAP}. The recent evidences in lead-free piezoelectric ceramics directly reveal that the ultrahigh longitudinal strain originates from the bending deformation of disk, driven by the oriented migration of oxygen vacancies under an external electric field\cite{tian2024defect,he2024ultra-MH,hou2024effects-CI}. While this phenomenon has been clarified, the necessary conditions for inducing electric-induced bending and the strong dependence of nominal strain on thickness remain unclear. To illustrate these questions, we conduct a theoretical derivation combined with molecular dynamics simulations to elucidate the underlying mechanisms of bending deformation. 

The constitutive equation for strain $\eta$ in a dielectric solid possessing piezoelectricity is given by
\begin{equation}
\eta _{ij}=S_{ijkl}{{\sigma}}_{kl}+d_{ijk}E_{k}
\end{equation}
where \textit{S} is the elastic compliance tensor, $\sigma$ is stress tensor, \textit{d} is piezoelectric strain tensor, and \textit{E} is the electric field. Since we consider the effect of electric field on mechanically free samples, the first term on right side, related to mechanical stress, can be disregarded. Then the strain is fully controlled by piezoelectric effect,
\begin{equation}
\eta _{ij}=d_{ijk}E_{k}
\end{equation}
We begin by analyzing the response of a piezoelectric beam under an external electric field, where the beam can be considered as a cross-section cross-sectional slice across the diameter of the piezoelectric disk. Assuming the electric field is applied only in the beam thickness direction (\textit{z}), since the electrodes are coated on upper and lower surfaces of the disk. The bending of beam should be viewed as the result of the inhomogeneous transverse strain, then the equation simplifies to 
\begin{equation}
\eta _{1}=d_{31}E_{3}
\end{equation}
where Vigot notation \textit{$\eta$ }\textsubscript{1} = \textit{$\eta$ }\textsubscript{11} and \textit{d}\textsubscript{31} = \textit{d}\textsubscript{311} are used for convenience. To inducing bending, the gradient of $\eta$\textsubscript{1} along \textit{z} direction should be non-zero,
\begin{equation}
\nabla _{z}\eta_1=E_{3}\nabla _{z}d_{31}+d_{31}\nabla _{z}E_{3}
\end{equation}
Suppose the dielectric constant $\epsilon$\textsubscript{33} of sample is independent on the thickness, so that the electric field across the piezoelectric disk is homogeneous due to the uniform potential applied by identically sized electrodes. This indicates that the second term related to \textit{$\nabla$}\textsubscript{z}\textit{E}\textsubscript{3} must be zero, so that only the inhomogeneous \textit{d}\textsubscript{31} allows to generate the necessary gradient of $\eta$\textsubscript{1} to induce bending.
\begin{figure}
        \centering
        \includegraphics[width=1\linewidth]{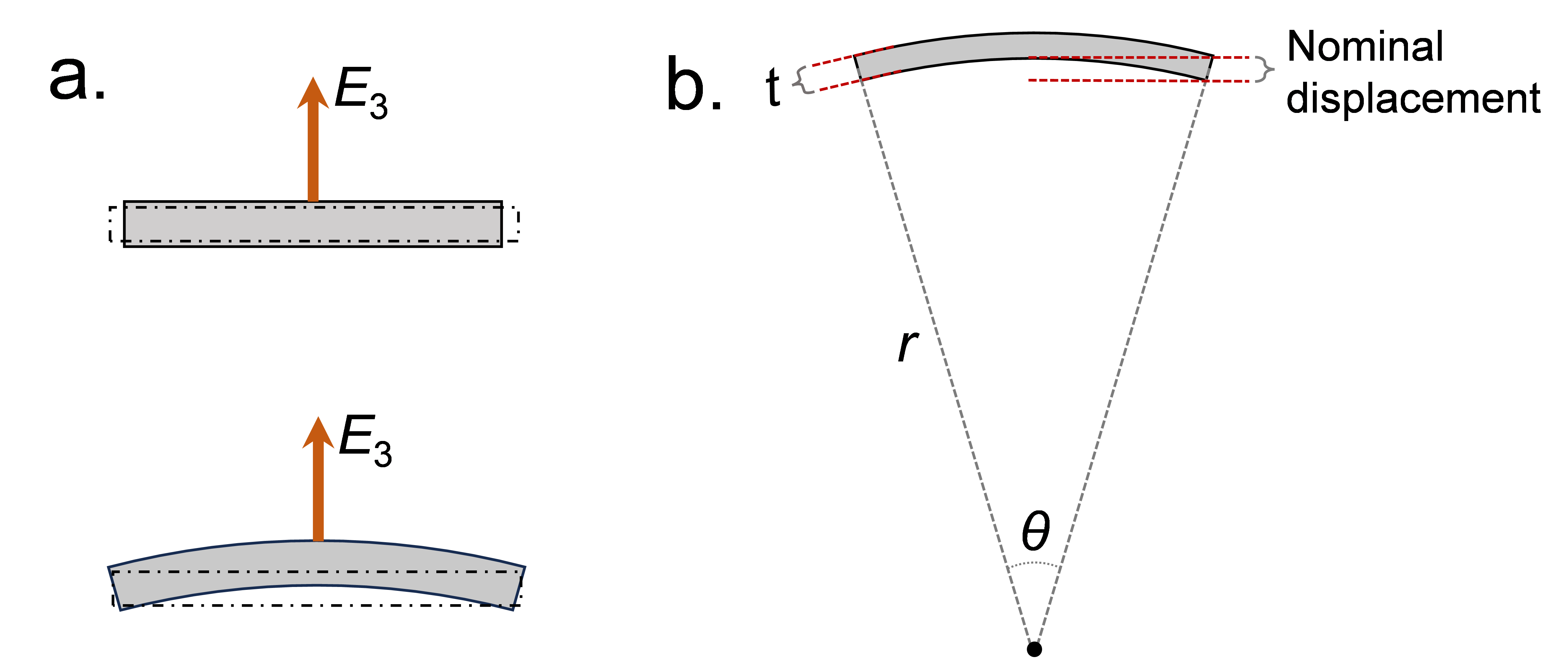}
        \caption{(a) Schematic diagram of beam with homogeneous and inhomogeneous \textit{d}\textsubscript{31} under an external field. (b) Schematic diagram of nominal strain induced by beam bending. }
        \label{fig l}
    \end{figure}
Next, we analyze how much nominal, or false, \textit{d}\textsubscript{33} is produced by bending due to gradient of \textit{d}\textsubscript{31}. Based on the assumptions of Euler-Bernoulli beam theory\cite{bauchau2009euler}, and by ignoring the longitudinal strain induced by \textit{d}\textsubscript{33} (which is typically on the order of 10\textsuperscript{-4}  $\sim$ 10\textsuperscript{-3}), we consider the bending of disk with a diameter of \textit{L}. Since the cross-section of the beam remains perpendicular to the neutral plane during bending, the bent beam with length \textit{L} can be treated as an arc, as shown in Fig 1. The following conditions must be satisfied for both sides of the arc
\begin{equation}
r\theta=L(1+d_{31}E_3)
\end{equation}
\begin{equation}
(r+t)\theta=L[1+(d_{31}+t\nabla_{z}d_{31})E_3]
\end{equation}
where the \textit{r} is the radius of lower-side arc, \textit{$\theta$} is the radian of arc, and \textit{t} is the thickness of beam. Solving for \textit{r} and $\theta$, we obtain
\begin{equation}
r=\frac{1+d_{31}E_{3}}{E_{3}\nabla_{z}d_{31}}\approx\frac{1}{E_{3}\nabla_{z}d_{31}}
\end{equation}
\begin{equation}
\theta=LE_{3}\nabla_{z}d_{31}
\end{equation}
The term of \textit{d}\textsubscript{31}\textit{E}\textsubscript{3}, representing the conventional transverse strain, is typically in the range of 10\textsuperscript{-4}  $\sim$  10\textsuperscript{-3} for perovskite ferroelectrics\cite{li2019-science, saito2004lead}, so that it can be reasonably ignored. Generally, the center of bent beam generates the mimetic longitudinal strain, as shown in Fig. 1b. In this case, the nominal longitudinal strain $\eta$\textsubscript{3,\textit{nom}} originated from the bending should be expressed as
\begin{equation}
\eta_{3,nom}=\frac{r(1-cos\frac{\theta}{2})}{t}
\end{equation}
After a simple calculation, the following relationship can be found
\begin{equation}
\eta_{3,nom}=\frac{1-cos({LE_{3}\nabla_{z}d_{31}}/{2})}{tE_{3}\nabla_{z}d_{31}}
\end{equation}
The equation indicates that the nominal strain is strongly dependent on the gradient of \textit{d}\textsubscript{31}, electric field \textit{E}\textsubscript{3}, the length \textit{L}, and the thickness \textit{t} of disk. To illustrate this relationship, we use a commonly employed disk diameter of 10 mm and an electric field of 5 kV/mm, setting gradient of \textit{d}\textsubscript{31} to values of 10\textsuperscript{-7}, 10\textsuperscript{-8}, 10\textsuperscript{-9}, and 10\textsuperscript{-10} V\textsuperscript{-1}, corresponding to 100, 10, 1, and 0.1 pm/V differences in \textit{d}\textsubscript{31} between the upper and lower side of disk with the thickness of 1 mm, and then the relationship between nominal strain and disk thickness can be found, as shown in Fig. 2a. We found that the nominal longitudinal strain resulting from the beam bending can be very large at a very thin sample thickness. For instance, a gradient of \textit{d}\textsubscript{31} at 10\textsuperscript{-8} V\textsuperscript{-1} can results in 1.25\% nominal strain at a thickness of 0.05 mm, which corresponds to a nominal \textit{d}\textsubscript{33} of 2500 pm/V that is much larger than the conventional \textit{d}\textsubscript{33} values.
\begin{figure}
    \centering
    \includegraphics[width=1\linewidth]{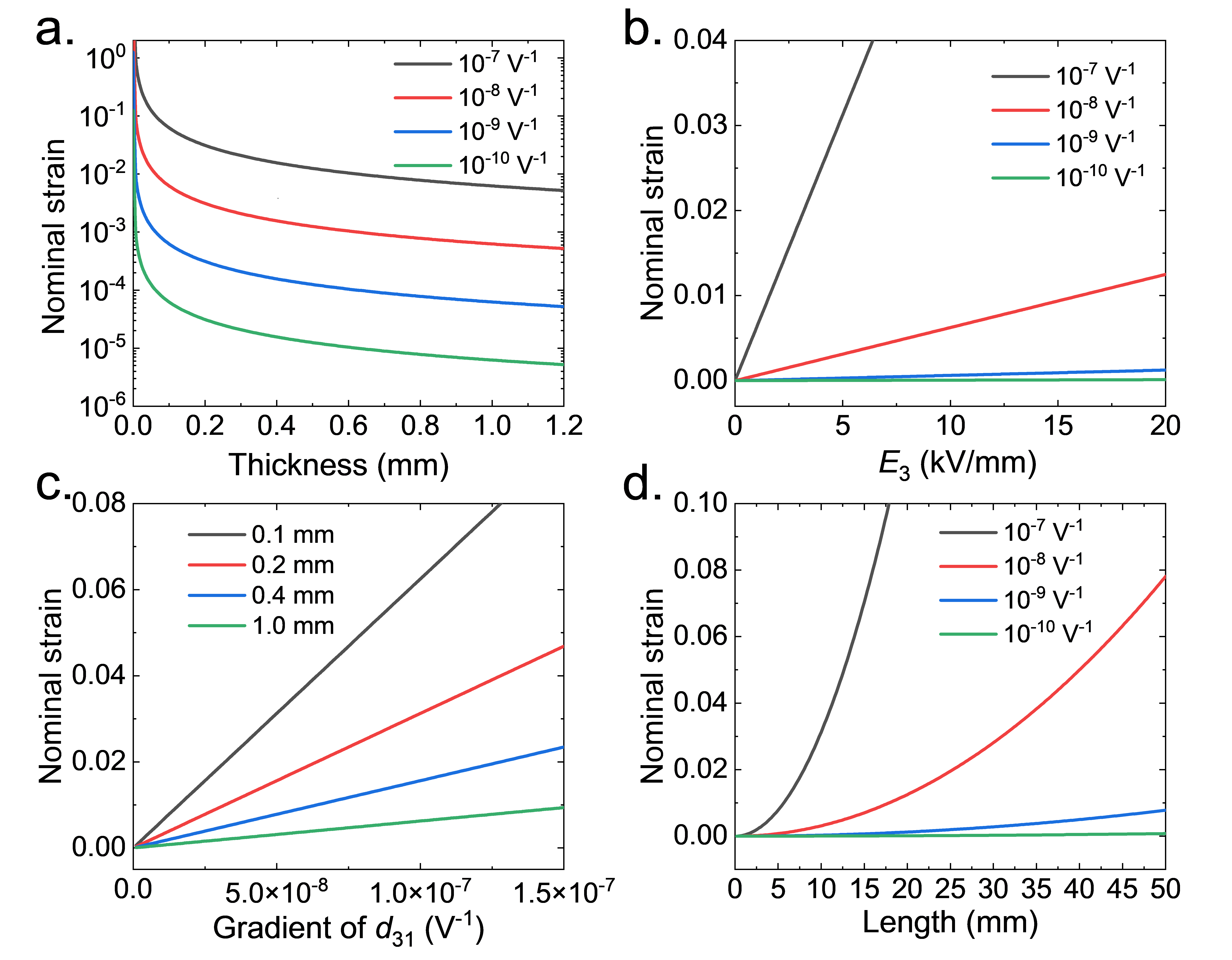}
    \caption{Nominal longitudinal strains as a function of (a) thickness, (b) electric field, (c) gradient of \textit{d}\textsubscript{31}, and (d) beam length. }
    \label{fig 2}
\end{figure}
We can also explore the relationship between nominal longitudinal strain and the electric field \textit{E}\textsubscript{3}. Using a length of 10 mm, a thickness of 0.2 mm, and gradients of \textit{d}\textsubscript{31} set as 10\textsuperscript{-7}, 10\textsuperscript{-8}, 10\textsuperscript{-9}, and 10\textsuperscript{-10} V\textsuperscript{-1}, the corresponding relationship is illustrated in Fig. 2b. In this case, the nominal longitudinal strain increases almost linearly with the change of \textit{E}\textsubscript{3}, indicating that the bending induced by gradient of \textit{d}\textsubscript{31} can closely mimic the longitudinal strain comes from \textit{d}\textsubscript{33}. Moreover, the nominal longitudinal strain is also directly proportional to the gradient of \textit{d}\textsubscript{31}, as shown in Fig. 2c, where the \textit{L} = 10 mm, \textit{E}\textsubscript{3} = 5 kV/mm, and \textit{t} = 0.1, 0.2, 0.4, 1 mm. The Fig. 2d indicates that increasing the beam length results in a rapid rise in nominal longitudinal strain, where the \textit{t} = 0.2 mm, \textit{E}\textsubscript{3} = 5 kV/mm, and $\nabla$\textsubscript{z}\textit{d}\textsubscript{31} = 10\textsuperscript{-7}, 10\textsuperscript{-8}, 10\textsuperscript{-9}, 10\textsuperscript{-10} V\textsuperscript{-1}. These relationships can be directly derived from Equation (9). Given that the term \textit{LE}\textsubscript{3}$\nabla$\textsubscript{z}\textit{d}\textsubscript{31}/2 is small, the term of cos(\textit{LE}\textsubscript{3}$\nabla$\textsubscript{z}\textit{d}\textsubscript{31}/2) can be expanded into a Taylor series at zero. In this expansion, the odd terms vanish, and only the even terms are retained. By discarding higher-order terms, it can be found that nominal longitudinal strain  $\eta$\textsubscript{3,\textit{nom}}  induced by beam bending is directly proportional to \textit{L}\textsuperscript{2}, \textit{E}\textsubscript{3}, and $\nabla$\textsubscript{z}\textit{d}\textsubscript{31}. Since bending creates a gradient of $E_{3}$, we also include the flexoelectric effect to account for curvature, however, it did not produce a noticeable difference\cite{Supplemental-Materials}.

The apparent strain is particularly high in ceramics containing defect dipoles, as reported in numerous experiments\cite{wang2024ultrahigh-AM,luo2023achieving-SA, park2022-CeO2}. In that case, the oxygen vacancies can be driven by an external electric field. This leads to an uneven distribution\cite{park2022-CeO2,tian2024defect}, where one side of the disk accumulates a large number of oxygen vacancies, while the opposite side maintains a more ordered polar lattice, as illustrated in Fig. 3a. The piezoelectricity on the side with oxygen vacancies is suppressed due to the presence of these defects. However, the decrease in piezoelectricity is not linear with the concentration of defects. Instead, piezoelectricity tends to drop rapidly at first, then stabilizes at a plateau as defect density increases. In this scenario, the difference of \textit{d}\textsubscript{31} between two sides of beam fixed value due to the accumulation of defects on one side. Then the gradient of \textit{d}\textsubscript{31} is thus approximately $\Delta$\textit{d}\textsubscript{31}/\textit{t}, and the nominal longitudinal strain can be expressed as 
\begin{equation}
\eta_{3,nom}=\frac{1-cos({LE_{3}\Delta d_{31}}/{2t})}{E_{3}\Delta d_{31}}
\end{equation}
We set \textit{L} = 10 mm,  $\Delta$\textit{d}\textsubscript{31} = 100, 10, 1, and 0.1 pm/V and apply an electric field of 5 kV/mm. The nominal longitudinal strains for different beam thicknesses are shown in Fig 3b. It is evident that the nominal longitudinal strain is strongly dependent on thickness. For example, when $\Delta$\textit{d}\textsubscript{31} = 10 pm/V, the  $\eta$\textsubscript{3,\textit{nom}} is 0.06\% at the thickness of 1 mm; However, when the thickness decreases to 0.1 mm, the $\eta$\textsubscript{3,\textit{nom}} jumps to 6.25\%, and further reducing the thickness to 0.05 mm, results in $\eta$\textsubscript{3,\textit{nom}} reaching an incredible 25\%. Additionally, for a given thickness and  $\Delta$\textit{d}\textsubscript{31}, $\eta$\textsubscript{3,\textit{nom}} continues to exhibit a linear relationship with changes in the applied electric field. 
\begin{figure}
    \centering
    \includegraphics[width=1\linewidth]{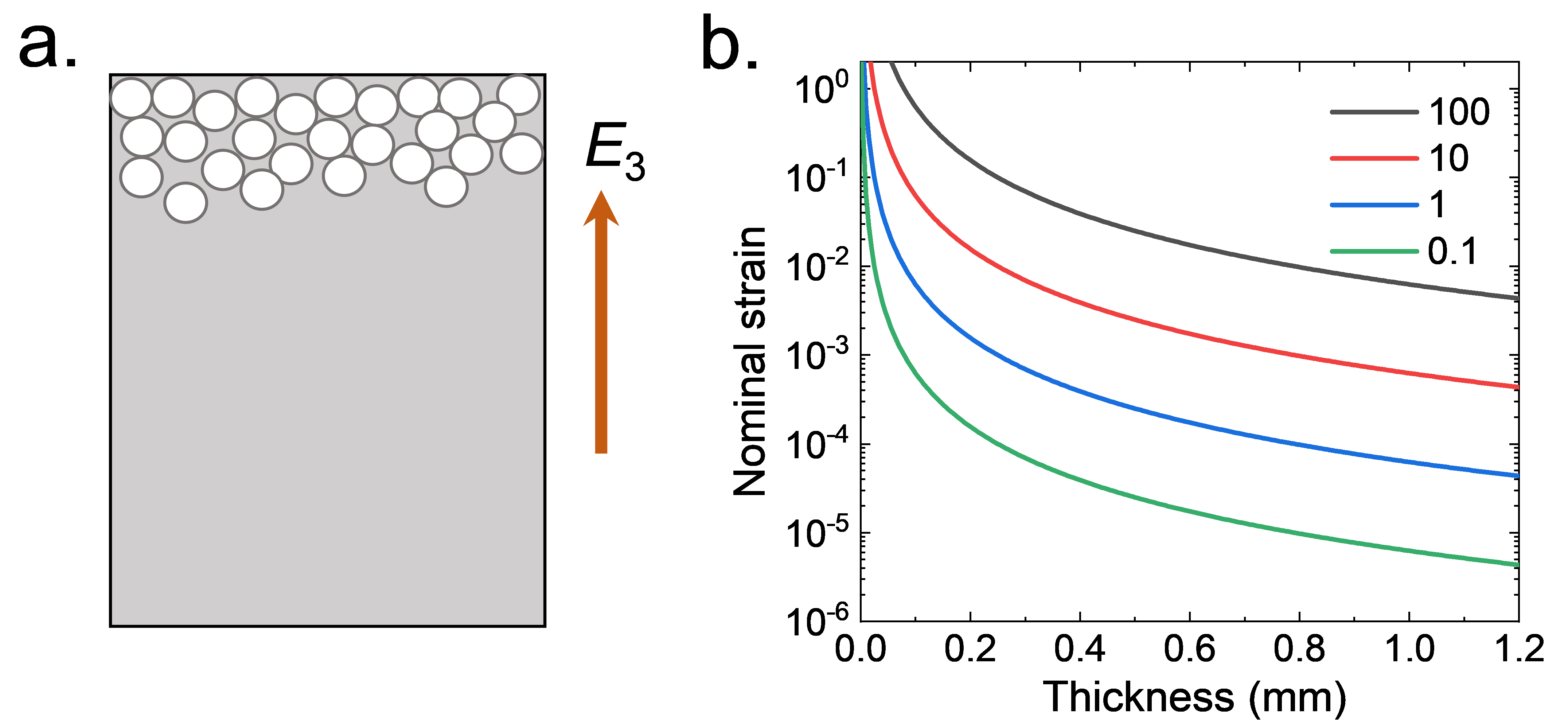}
    \caption{(a) Schematic diagram of defect gathering under the application of electric field. (b) Nominal longitudinal strain as a function of thickness at different $\Delta$\textit{d}\textsubscript{31}.}
    \label{fig:3}
\end{figure}

To assess the intrinsic influence of defect dipole on the piezoelectric properties of orthorhombic KNbO\textsubscript{3}, we conduct molecular dynamics simulations based on the on-the-fly machine learning force fields using Vienna ab initio simulation package (VASP)\cite{kresse1994ab-PRB,kresse1996efficient-PRB}. For comparison, we set potassium deficiency at 0\%, 2.08\%, and 4.17\% in a 12 × 8 × 8 orthorhombic supercell with \textit{Amm}2 group, corresponding to oxygen vacancy concentrations of 0\%, 0.35\%, and 0.69\%, respectively. Since phase transition temperatures are typically underestimated in simulations, we performed the simulations in the \textit{NPT }ensemble for 100 ps with a 1 fs timestep at 80 K, maintaining the orthorhombic symmetry. The piezoelectric constant was calculated using a time-averaging method based on the equation\cite{garcia1998electromechanical-APL}
\begin{equation}
d_{ij}=\frac{1}{k_{B}T}\langle\Delta M_{i}\Delta \eta_{j}\rangle
\end{equation}
where \textit{k\textsubscript{B}} is the Boltzmann constant, \textit{T} is the temperature, \textit{M\textsubscript{i}} is the \textit{i}th total dipole moment of supercell, and $\eta$\textsubscript{\textit{j}} is the \textit{j}th element of strain tensor. The piezoelectric strain constants obtained from our simulations are listed in Table 1. Notably, the piezoelectric constants associated with polarization stretching (ie., \textit{d}\textsubscript{31}, \textit{d}\textsubscript{32}, and \textit{d}\textsubscript{33}) remained largely unchanged after the introduction of vacancies. However, the \textit{d}\textsubscript{24} and \textit{d}\textsubscript{15} are significantly reduced, indicating that polarization rotation is strongly suppressed by defect dipoles. At first glance, one might assume that the defects do not strongly affect the \textit{d}\textsubscript{31} of KNbO\textsubscript{3} ceramics. However, the polycrystalline ceramics is a assemble of grains with various orientations, meaning that the \textit{z} axis of any given grain in the Cartesian coordinate system \{\textit{x}, \textit{y}, \textit{z}\} does not always align with the out-plane direction of ceramic disk. Therefore, the \textit{d}\textsubscript{31} of piezoelectric ceramics should be an average of each grain’s out-plane contribution. The piezoelectric constants of KNbO\textsubscript{3} ceramics can be estimated by an orientational average method\cite{Supplemental-Materials,tan2019intrinsic}. For an orthorhombic perovskite ferroelectric, the transverse piezoelectric constant of ceramics, \textit{d}\textsubscript{31,\textit{c}} , has follow relation\cite{tan2019intrinsic, peng2019tunable}
\begin{equation}
\begin{split}
d_{31,c}&\approx 0.4128d_{31}+0.4272d_{32}+0.0722d_{33}\\
&-0.0433d_{15}-0.0289d_{24}
\end{split}
\end{equation}
Using this approach, the transverse piezoelectric constants of KNbO\textsubscript{3} ceramics are calculated to be -13.9, -9.4, and -7.6 pC/N for oxygen vacancy levels of 0\%, 0.35\%, and 0.69\%, respectively. The difference in \textit{d}\textsubscript{31} between 0\% and 0.69\% oxygen-deficient system reaches 6.3 pC/N, which also indicates the that the disk bulges on one side of the accumulated oxygen vacancy, agreeing with the recent experiment observation\cite{tian2024defect}. It is important to note that the piezoelectricity (e.i., \textit{d}\textsubscript{31}, \textit{d}\textsubscript{33}, and \textit{d}\textsubscript{15}) of ceramics are significantly affected by the shear piezoelectric constants of its crystal, and the shear piezoelectricity constants of ceramics can be extremely enhanced by engineering coexisting multiple ferroelectric phases. Hence, the vacancies in high-performance composition will lead to higher $\Delta$\textit{d}\textsubscript{31}. An obvious example is that the temperature-dependent nominal longitudinal strain in K\textsubscript{0.5}Na\textsubscript{0.5}NbO\textsubscript{3} system reported by Tian et al.\cite{tian2024defect}, which reaches a maximum of 18\% around 210 ℃, near the orthorhombic and tetragonal phase boundary. This implies that, at the room temperature, defect-induced $\Delta$\textit{d}\textsubscript{31} for K\textsubscript{0.5}Na\textsubscript{0.5}NbO\textsubscript{3 }system could exceed 10 pC/N, leading to very large nominal longitudinal strains in thin ceramic disk, with $\eta$\textsubscript{3,nom} exceeding 6\% at a thickness of 0.1 mm, as shown in Fig 3b. The 0.67\% difference in oxygen vacancy concentration is plausible due to the directional migration under an external electric field, as several experiments have shown oxygen vacancy migration\cite{park2022-CeO2,tian2024defect}. Therefore, in high-performance piezoelectric materials, variations in oxygen vacancy concentrations within perovskite ferroelectrics can lead to differences of several to hundreds of pC/N in \textit{d}\textsubscript{31} of ceramics, thereby promoting beam bending and resulting in large nominal longitudinal strains.

\begin{table}
\centering
\caption{Obtained piezoelectric strain constants \textit{d\textsubscript{ij}} of KNbO\textsubscript{3} at 80 K, in the unit of pC/N. The subscript $c$ represents the properties of the ceramic.}
\label{tab 1}
\begin{tabular}{l >{\raggedright\arraybackslash}p{0.09\linewidth}>{\raggedright\arraybackslash}p{0.09\linewidth}>{\raggedright\arraybackslash}p{0.09\linewidth}>{\raggedright\arraybackslash}p{0.09\linewidth}>{\raggedright\arraybackslash}p{0.09\linewidth}>{\raggedright\arraybackslash}p{0.09\linewidth}>{\raggedright\arraybackslash}p{0.09\linewidth}>{\raggedright\arraybackslash}p{0.09\linewidth}}
\hline
 \makecell{Oxygen \\ vacancy}& 
 \textit{d}\textsubscript{31} & \textit{d}\textsubscript{32} & \textit{d}\textsubscript{33} & \textit{d}\textsubscript{24} & \textit{d}\textsubscript{15} & \textit{d}\textsubscript{31,\textit{c}} & \textit{d}\textsubscript{33,\textit{c}} & \textit{d}\textsubscript{15,\textit{c}} \\
\hline
0\% & -24.5 & 26.1 & 33.5 & 472.1 & 85.0 & -13.9 & 59.7 & 239.5 \\
0.35\% & -24.9 & 25.8 & 33.8 & 323.8 & 74.7 & -9.4 & 50.5 & 172.0 \\
0.69\% & -23.5 & 24.0 & 32.3 & 274.3 & 60.2 & -7.6 & 45.2 & 144.7 \\
\hline

\end{tabular}

\end{table}
Building on the above results, it is speculated that some recently reported large longitudinal piezoelectric strains are also induced by the bending, where the $\Delta$\textit{d}\textsubscript{31} is not only induced by oxygen vacancies, but also by other factors. The previous work\cite{hu2020ultra-SA} reported an ultra-large electric-induced strain in in the as-grown [001]\textsubscript{c}-oriented K\textit{\textsubscript{x}}Na\textsubscript{1-\textit{x}}NbO\textsubscript{3} crystal with an engineered compositional gradient. We noted that the large electric-induced strain was only observed in the specific composition of K\textsubscript{0.43}Na\textsubscript{0.57}NbO\textsubscript{3}, while other compositions, such as K\textsubscript{0.35}Na\textsubscript{0.65}NbO\textsubscript{3} and K\textsubscript{0.44}Na\textsubscript{0.56}NbO\textsubscript{3}, can only exhibited strains below 0.06\%. The top-seeded solution growth method used in the preparation of these crystals often results in a compositional gradient, which means the compositions of top and bottom surfaces of sample is subtly different. In most cases, these compositional differences do not significantly affect piezoelectricity, given that the crystal symmetry remains the same in K\textit{\textsubscript{x}}Na\textsubscript{1-\textit{x}}NbO\textsubscript{3} system. However, near the K : Na = 0.43 : 0.57 composition ratio, the phase diagram\cite{baker2009comprehensive-APL} reveals a phase transition associated with the oxygen octahedron rotation at room temperature. Around this critical boundary, higher potassium content expands the unit cell, reducing compressive stress on the O-Nb-O chain, stabilizing the orthorhombic phase without oxygen octahedron rotation, denoted as $a_{+}^{0}$$a_{+}^{0}$$c_{0}^{0}$. In contrast, the lower potassium content will shrinks the unit cell, subjecting the O-Nb-O chain to significant compressive stress. This leads to in-phase tilting of the oxygen octahedron along the non-polar direction to ease the Nb-O bond, denoted as  $a_{+}^{0}$$a_{+}^{0}$$c_{0}^{+}$. Despite these two distinct structures keep direction of the polarization vector, [110]\textsubscript{c}, the tilting of oxygen octahedron affects the local environment of Nb atom. This creates a high energy barrier that hinders polarization rotation, significantly decreasing the shear piezoelectric constants\cite{yan2017crucial-ActaMater}. In other words, the shear piezoelectric constants are highly sensitive near the K : Na = 0.43 : 0.57 boundary. As a result, the difference in shear piezoelectric constants between the top and bottom of a [001]\textsubscript{c}-oriented K\textsubscript{0.43}Na\textsubscript{0.57}NbO\textsubscript{3} crystal plate can be substantial enough to induce bending, then this bending effect may give rise to an apparent high strain that is observed experimentally. 

While writing this paper, we also noted the recent report claiming that high longitudinal strain of 2.5\% was achieved in NaNbO\textsubscript{3} film grown on the SrTiO\textsubscript{3} substrates\cite{lin2024ultrahigh-NN}. The high strain is attributed to the competing antiferroelectric and ferroelectric orders within this film. However, we found that this explanation may not fully account for the observed strain. Despite the phase transition from the \textit{Pbma} to \textit{R3c} does induce the a change in lattice parameter, from 3.898 to 3.919 Å, the maximum strain produced from this change is only around 0.54\%. Taking into account the phase ratio and crystal orientation, the experimentally obtained strain should be even smaller than this, falling well short of the reported 2.5\%. The true origin of high longitudinal strain may still be related to bending effects in the film. The phase transition from \textit{Pbma} to \textit{R3c} induce the transverse expansion, which corresponds to an effective positive \textit{d}\textsubscript{31} at the top surface of NaNbO\textsubscript{3} film under the mechanically free conditions. However, the bottom of NaNbO\textsubscript{3} film is constrained by the SrTiO\textsubscript{3} substrates, which prevents transverse strain and effectively keeps \textit{d}\textsubscript{31} at zero. This creates a non-zero gradient of \textit{d}\textsubscript{31} across the film thickness, making it plausible that the bending of the film is responsible for the observed longitudinal strain. In this scenario, the film would bend such that the maximum longitudinal strain occurs at the center, with the strain decreasing towards the edges, which would experience zero strain. The result is consistent with their experimental strain characterization\cite{lin2024ultrahigh-NN}. Furthermore, the reported dependence of longitudinal strain on frequency in NaNbO\textsubscript{3} films could also be attributed to the relaxation dynamics of this bending. At lower frequencies, the film has more time to relax, leading to higher observed strains, while at higher frequencies, the film is unable to fully relax, resulting in lower strains.

The gradients of defect concentration, composition, and stress can all cause an inhomogeneous distribution of \textit{d}\textsubscript{31} in the thickness direction. It is important to recognize that oxygen vacancy induced by the volatilization of A-site elements (e.g., K, Na, Pb, Bi, etc.) during the sintering process is inevitable. As a result, electric-induced bending may be more common in perovskite ferroelectrics than previously acknowledged. To avoid attributing mimetic strains induced by bending to actual electric-induced strains, the following approaches are recommended to verify true electric-induced strain: 

i. Performing local measurements. Use a pointed clip to clamp the sample locally rather than laying it flat on a platform. This setup allows for direct detection of thickness changes while excluding the effects of bending. 

ii. Checking the effect of sample size on the longitudinal strain. Equation (10) indicate the nominal longitudinal strain caused by bending is strongly dependent on the length and thickness of the beam. In contrast, true electric-induced strain should remain independent of sample size.

iii. Using suitable geometric design. Samples with shorter lengths and thicker cross-sections will minimize the bending effect. For example, using the equation (9), even with a high gradient of $\nabla$\textsubscript{z}\textit{d}\textsubscript{31} = 10\textsuperscript{-7} V\textsuperscript{-1}, a sample with \textit{L} = 1 mm, \textit{t} = 1 mm, and an applied field \textit{E} = 5 kV/mm would only result in a bending-induced nominal strain of 0.006\%, much smaller than the typical longitudinal strain ($\sim$ 0.1\%) in perovskite ferroelectrics.

While this bending phenomenon complicates the interpretation of piezoelectric experiments, it presents a valuable opportunity to enhance piezoelectric bending actuators. This piezoelectric bending effect closely resembles so-called inverse flexoelectricity (curvature induced by a uniform electric field)\cite{wen2021inverse-PRApplied,zubko2013flexoelectric} but can achieve significant curvature on the millimeter scale. Conventional bimorph actuators typically consist of two piezoelectric plates or two plates with an elastic shim bonded together. However, the bonding layer in the latter design increases hysteresis, degrades displacement performance, and leads to delamination issues\cite{uchino2015glory}. Moreover, the fabrication process, which involves cutting, polishing, electroding, and bonding, is labor-intensive and costly\cite{uchino2015glory}. Thus, a monolithic piezoelectric bending actuator eliminating the need for bonding is a very attractive alternative structure.

In summary, we reveal that bending induced by a non-zero gradient of \textit{d}\textsubscript{31} in thickness direction in the thickness direction is not only possible but can lead to abnormally high nominal longitudinal strains in samples with thin thickness and long length. Our simulations further demonstrate that vacancies reduce the shear piezoelectric constants, influencing the overall \textit{d}\textsubscript{31} in ceramics. The gradients in defect concentration, composition, and stress can all contribute to the inhomogeneous distribution of \textit{d}\textsubscript{31}, creating challenges in interpreting experimental data accurately. We have proposed several experimental strategies to differentiate between true electric-induced strain and bending-induced nominal strain. Additionally, the piezoelectric bending effect offers a practical opportunity to create piezoelectric bending actuators that are simple and cost-effective, paving the way for innovative applications that leverage these amplified piezoelectric responses.

~

\begin{acknowledgments}
This work was supported by the National Natural Science Foundation of China (grants 12404110, 51932010, 52032007, 52202294), the Sichuan Science and Technology Program (2023YFG0042), the Natural Science Foundation of Sichuan Province (2024NSFSC1383), and the Fundamental Research Funds for the Central Universities.  The authors are also grateful for the useful discussion with Hao-Cheng Thong.
\end{acknowledgments}

\bibliography{APS-Bending}
\end{document}


\preprint{APS/123-QED}

\title{Supplemental Materials for \\Transverse Bending Mimicry of Longitudinal Piezoelectricity}

\author{Zhi Tan}
 \email{tanzhi0838@scu.edu.cn}
 \affiliation{College of materials science and engineering, Sichuan University, Chengdu 610065, China.}

\author{Xiang Lv}
\affiliation{College of materials science and engineering, Sichuan University, Chengdu 610065, China.}
 
\author{Jie Xing}%
 \email{xingjie@scu.edu.cn}
  \affiliation{College of materials science and engineering, Sichuan University, Chengdu 610065, China.}
  
\author{Shaoxiong Xie}
\affiliation{Department of Materials Science and Engineering, Friedrich-Alexander-Universität Erlangen-Nürnberg (FAU), Erlangen 91058, Germany}

\author{Hui Zhang}
\email{zhanghui@xidian.edu.cn}
\affiliation{
 Shaanxi Key Laboratory of High-Orbits-Electron Materials and Protection Technology for Aerospace, School of Advanced Materials and Nanotechnology, Xidian University; Xi'an 710126, China 
}%

 \author{Jianguo Zhu}%
 \email{nic0400@scu.edu.cn}
\affiliation{%
 College of materials science and engineering, Sichuan University, Chengdu 610065, China\\
}%

\date{\today}

\begin{abstract}

. 

\end{abstract}

\maketitle

\renewcommand\thefigure{\thesection S\arabic{figure}}    

\subsection{\label{sec:level1}Computational Method}
The first-principle calculations in the present work are based on density functional theory (DFT) as implemented in Vienna Ab initio Simulation Package (VASP)\cite{kresse1994ab-PRB,kresse1996efficient-PRB},. The Perdew-Burke-Ernzerhof revised for solids (PBEsol) \cite{perdew2008restoring-PBEsol} exchange-correlation functional with projector-augmented wave (PAW)\cite{blochl1994projector} is employed. The K 3\textit{s}\textsuperscript{2}3\textit{p}\textsuperscript{6}4\textit{s}\textsuperscript{1}, Nb 4\textit{p}\textsuperscript{6}5\textit{s}\textsuperscript{2}4\textit{d}\textsuperscript{3}, and O 2\textit{s}\textsuperscript{2}2\textit{p}\textsuperscript{4} orbitals are treated as valence electrons for calculations. A plane wave expansion with a cut-off energy of 520 eV is used to represent the wave function. $\Gamma$-centered \textit{k}-point meshes with a grid of spacing 0.04 × 2$\pi$ \AA\textsuperscript{-1} are selected for Brillouin zone sampling. The Kohn-Sham orbitals are iteratively updated in the self-consistency cycle until an energy convergence of 10\textsuperscript{-5} eV is obtained, and geometry optimizations proceed until the residual atomic forces are below 0.01 eV \AA \textsuperscript{-1}. The on-the-fly machine learning force fields (MLFF)\cite{jinnouchi2019fly,jinnouchi2019phase} are firstly trained using a 3 × 3 × 3 perovskite supercell without defect dipoles to conduct the \textit{ab initio} molecular dynamics (MD) in VASP. Subsequent training was done on a same supercell with defect dipoles. These simulations are executed in the \textit{NPT} ensemble for 20 ps with the time step of 1 fs and energy convergence criterion of 10\textsuperscript{-4} eV, at the temperature of 20, 80, 150, 250, 350, and 500 K, respectively. A total of 1061 configurations were collected for the training dataset. The root mean squared errors (RMSEs) of energies, forces, and stress of obtained MLFF predictions with respect to \textit{ab initio} results for the training data are 1.12 × 10\textsuperscript{-3} eV per atom, 5.47 × 10\textsuperscript{-2} eV \AA\textsuperscript{-1}, and 0.791 kB, respectively. Using the generated MLFF, MD simulations are performed on a 12 × 8 × 8 orthorhombic supercell containing 7680 atoms, with a total times of 100 ps and time step of 1 fs at a constant temperatures of 80 K. The temperature of 80 K allows that the structure remains in the orthorhombic phase in simulations, ensuring consistency with the experimental  symmetry observed at room temperature.

\subsection{\label{sec:level1}\textbf{Curvature Derivation by Including Flexoelectric Effect}
}
We also can consider the beam bending in term of stress based on Euler-Bernoulli beam theory\cite{bauchau2009euler}. In there, we include the flexoelectric effect and the stress $\sigma$ on the beam's cross section with infinitesimal length can be written as
\begin{equation}
\sigma _{ij}=C_{ijkl}{{\eta}}_{kl}-e_{kij}E_{k}+\mu_{ijkl}\frac{\partial E_{k}}{\partial x_{l}} \tag{S1}
\end{equation}
where $\sigma$ is stress tensor, $C$ is elastic stiffness tensor, $e$ is piezoelectric stress tensor, $E$ is the electric field, and $\mu$ is the flexoelectric tensor. Because the beam cross-sections that are symmetrical about a plane perpendicular to the neutral plane, and the assumptions that the cross-section is infinitely rigid in its own plane and remains plane after deformation, indicating that \textit{C\textsubscript{ij}} = 0 when \textit{i} $\neq$ \textit{j} and only the stress $\sigma$\textsubscript{1} (Vigot notation) needs to be taken into account. In this case, the stress $\sigma$\textsubscript{1} is dependent on the thickness 
\begin{equation}
\sigma _{1,z}=C_{11}{{\eta}}_{1,z}-e_{31,z}E_{3}+\mu_{15}\frac{\partial E_{3}}{\partial z} \tag{S2}
\end{equation}
here, we use the \textit{z} to define the distance from the neutral plane, as shown in Fig S1, and $\mu$\textsubscript{15} = $\mu$\textsubscript{1133}. As the assumptions that the beam cross-section is perpendicular to the central axis, the strain $\eta$\textsubscript{1} in a section with infinitesimal length can be 
\begin{equation}
dx=rd\theta \tag{S3}
\end{equation}
\begin{equation}
\eta_{1,z}=\eta_{1,0}+\frac{(r+z)d\theta-dx}{dx}=\eta_{1,0}+\frac{z}{r} \tag{S4}
\end{equation}
Then we consider the gradient of \textit{e}\textsubscript{31} is constant, having
\begin{equation}
e_{31,z}=e_{31,0}+z\nabla_{z}e_{31} \tag{S5}
\end{equation}
And the gradient of \textit{E}\textsubscript{3} in the thickness of beam is given by
\begin{equation}
\frac{\partial E_{3}}{\partial z}=-\frac{rE_{3,0}}{(r+z)^2}\tag{S6}
\end{equation}
in where \textit{E}\textsubscript{3,0} represents the electric field at the $z=0$. We can see that if the beam does not bend, the \textit{r} should be infinity, then the gradient of \textit{E}\textsubscript{3} is zero. The stress is expressed as
\begin{align}
\sigma_{1,z}=&C_{11}\eta_{1,0}+\frac{C_{11}z}{r} -e_{31,0}E_{3}-zE_{3}\nabla_{z}e_{31} \nonumber\\
&-\mu_{15}\frac{rE_{3,0}}{(r+z)^2}\tag{S7}
\end{align}
At the \textit{z} = 0, the mechanically free condition allows
\begin{equation}
    \sigma_{1,0}=C_{11}\eta_{1,0}-e_{31,0}-\mu_{15}\frac{E_{3,0}}{r}=0 \tag{S8}
\end{equation}
Then the Equation (S7) can be written as
\begin{align}
\sigma_{1,z}&=\frac{C_{11}z}{r} -zE_{3}\nabla_{z}e_{31}+\mu_{15}\frac{z(2r+z)E_{3,0}}{r(r+z)^2}\nonumber\\ &\approx\frac{C_{11}z}{r} -zE_{3}\nabla_{z}e_{31}+\mu_{15}\frac{2zE_{3}}{r^{2}}
\tag{S9}
\end{align}
We consider the radian $\theta$ and thickness of beam are small, so that having \textit{z} $\ll$ \textit{r} and $E_{3}\approx E_{3,0}\approx U/t$, \textit{U} is the voltage on the disk, and \textit{t} is the thickness. The equilibrium gives $\sigma$\textsubscript{1,z} = 0, then we can solve $\frac{1}{r}$ 
\begin{align}
    \frac{1}{r}=\frac{-C_{11}\pm\sqrt{C_{11}^2+8\mu_{15}E_{3}^2\nabla_{z}e_{31}}}{4\mu_{15}E_{3}} \tag{S10}
\end{align}
The flexoelectric coefficient $\mu$\textsubscript{15 }is in the order of 10\textsuperscript{-9} C/m, which is far smaller than the \textit{C}\textsubscript{11} (in the order of 10\textsuperscript{11} Pa). Using Taylor series, the second term can be expanded at $\mu_{15}=0$ and discarding the abnormal value and high-order terms,  then 
\begin{align}
    \frac{1}{r}&\approx\frac{-C_{11}+(C_{11}+\frac{4E_{3}^2\nabla_{z}e_{31}}{C_{11}}\mu_{15}-\frac{8E_{3}^4(\nabla_{z}e_{31})^2}{C_{11}^3}\mu_{15}^2)}{4\mu_{15}E_{3}} \nonumber\\
    &=\frac{E_{3}\nabla_{z}e_{31}}{C_{11}}-\frac{2E_{3}^3(\nabla_{z}e_{31})^2\mu_{15}}{C_{11}^3}\tag{S11}
\end{align}
Recalling Equation (S3), the $\theta$ can be solved by
\begin{align}
\theta&=\int _{0}^{L} [\frac{E_{3}\nabla_{z}e_{31}}{C_{11}}-\frac{2E_{3}^3(\nabla_{z}e_{31})^2\mu_{15}}{C_{11}^3}]dx \nonumber\\ 
&=LE_{3}\nabla_{z}d_{31}-\frac{2LE_{3}^3(\nabla_zd_{31})^2\mu_{15}}{C_{11}}  \tag{S12}
\end{align}
where \textit{L} is the length of beam and $\nabla$\textit{\textsubscript{z}}\textit{d}\textsubscript{31} = $\nabla$\textit{\textsubscript{z}}\textit{e}\textsubscript{31}/\textit{C}\textsubscript{11} when \textit{C}\textsubscript{12} = \textit{C}\textsubscript{13} = 0. The negative sign indicates that the flexoelectric effect not only does not promote bending deformation, but also hinders it. Anyway, the second term as a whole is very small (in the order of $10^{-13}\sim 10^{-15}$), suggestting that flexoelectric effect is not important to bending deformation in piezoceramic disks at this case. In addition,  although start point is the stress in here, the $\theta$ is equal to the Equation (8)  at $\mu_{15}=0$.

\begin{figure}
        \centering
        \includegraphics[width=0.7\linewidth]{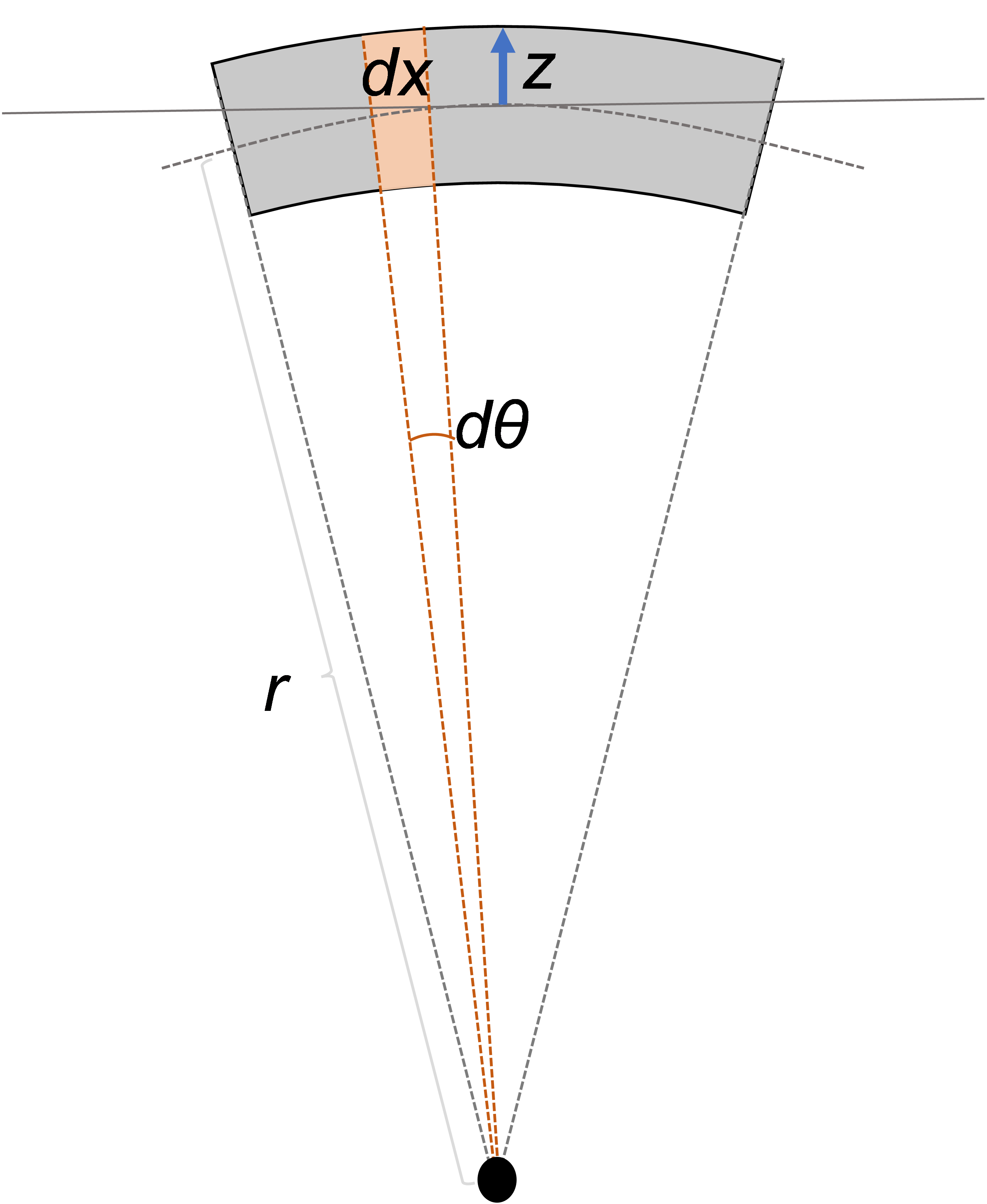}
       \caption{Schematic diagram of a bent beam: the fibers form concentric arcs, the top fibers are stretched and bottom fibers compressed.}
        \label{Figure S1}
    \end{figure}
\begin{figure*}
    \centering
    \includegraphics[width=1\linewidth]{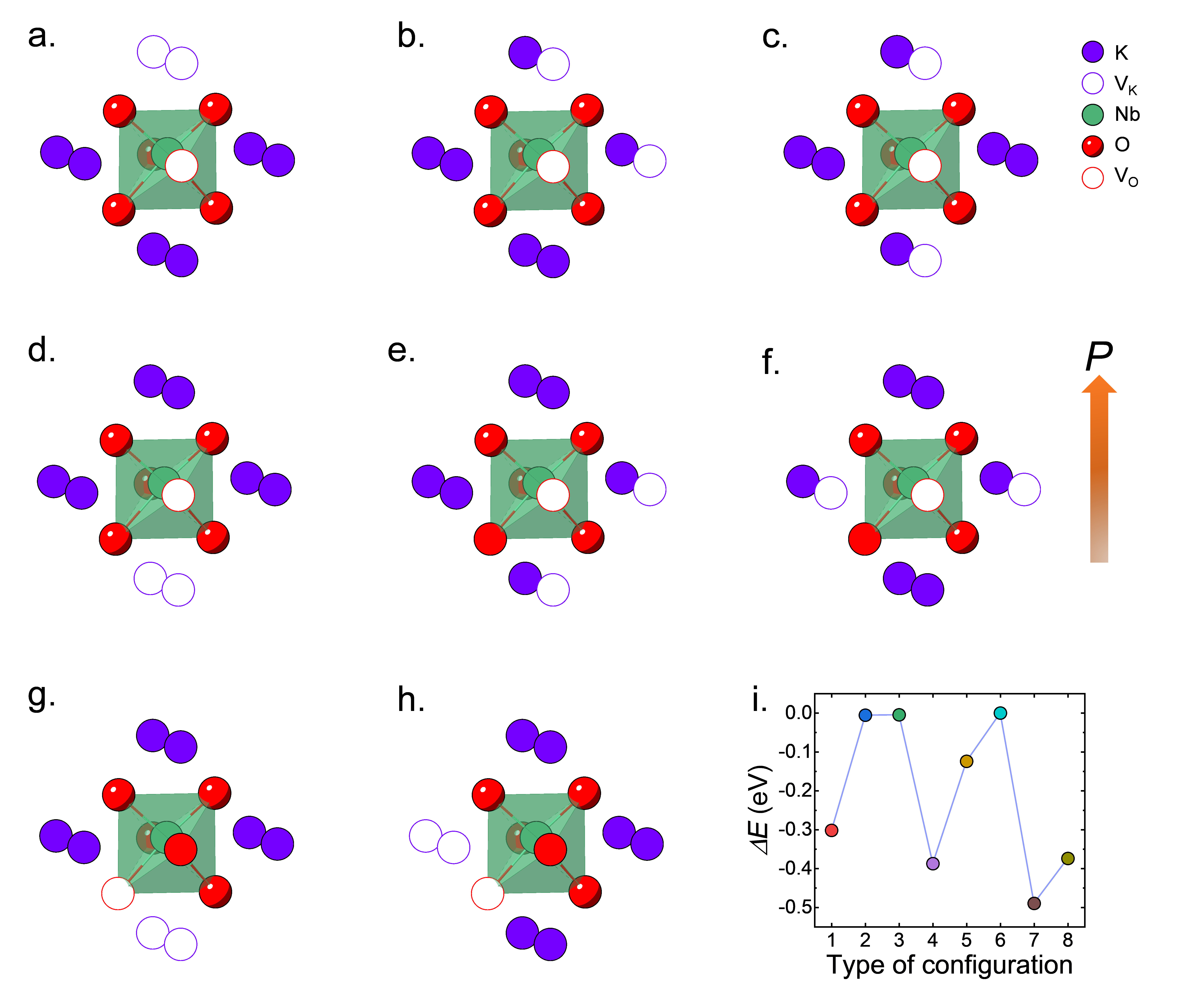}
    \caption{(a-h) The configurations of supercell contained defect dipole from type 1 to 8. (i) The total energy difference in these configurarions. V\textsubscript{K} and V\textsubscript{O} represent the vacancy of K and O, respectively.  }
    \label{Figure S2}
\end{figure*}

\subsection{\label{sec:level1}\textbf{\textbf{In the Case of Inhomogeneous Dielectric Constant}}}
One may think that the inhomogeneous case should be taken into account, because the defects concentration, gradient of composition, etc. usually induce the inhomogeneous dielectric constant $\epsilon$\textsubscript{33} in ceramics sample. However, the electric displacement should be continuous and unchanging in the sample. Let us recall the Equation (3), then 
\begin{equation}
\eta_{1}=d_{31}E_3=g_{31}D_{3} \tag{S13}
\end{equation}
Where \textit{g}\textsubscript{31} is the transverse piezoelectric voltage constant, and \textit{D}\textsubscript{3} is the electric displacement in \textit{z}-direction, \textit{d}\textsubscript{31} = \textit{g}\textsubscript{31}$\epsilon$\textsubscript{33}, and \textit{E}\textsubscript{3} = \textit{D}\textsubscript{3}/$\epsilon$\textsubscript{33}. Since the \textit{D}\textsubscript{3} is continuous in the sample, the gradient of \textit{D}\textsubscript{3} in \textit{z}-direction must be zero in a parallel-plate capacitor, then
\begin{equation}
\nabla_{z}\eta_{1}=D_{3}\nabla_{z}g_{31}+g_{31}\nabla_{z}D_{3} \tag{S14}
\end{equation}
The second term should be zero due to the continous \textit{D}\textsubscript{3}. The equation (S14) indicates that the gradient of $\eta$\textsubscript{1} is determined by $\nabla$\textsubscript{z}\textit{g}\textsubscript{31}. Considering the same piezoelectric beam, then solving the radius \textit{r }of lower-side arc, the radian \textit{$\theta$}, and nominal strain
\begin{equation}
r=\frac{1+g_{31}D_{3}}{D_{3}\nabla_{z}g_{31}} \approx\frac{1}{D_{3}\nabla_{z}g_{31}}\tag{S15}
\end{equation}
\begin{equation}
\theta=LD_{3}\nabla_{z}g_{31}\tag{S16}
\end{equation}
\begin{equation}
\eta_{3,nom}=\frac{1-cos(LD_{3}\nabla_{z}g_{31})/2}{tD_{3}\nabla_{z}g_{31}}\tag{S17}
\end{equation}
The equation (S17) is similar to the equation (10), hence, exhibiting similar characteristics for changing thickness \textit{t}, length \textit{L}, electric displacement \textit{D}\textsubscript{3}, and the gradient of \textit{g}\textsubscript{31}.

\subsection{\label{sec:level1}\textbf{\textbf{\textbf{\textbf{Determining the Type of Defect Dipole} }
}}}
We consider the eight types of defect dipole in a 3 × 2 × 2 ferroelectric orthorhombic KNbO\textsubscript{3} supercell. After the sufficient ionic relaxation, we compare their total energy to find the most possible configuration in orthorhombic KNbO\textsubscript{3}. We find the type 7 is the most stable configuration due to the lowest total energy, as shown in Figure S2i. A little later, the configuration of type 7 is chosen to conduct the training of MLFF.

\subsection{\label{sec:level1}\textbf{\textbf{\textbf{Inhibited Polarization Fluctuation}
}}}
The distribution of polarization differences from our simulations is shown in Figure S3. The results indicate that the distributions of $\Delta$\textit{P}\textsubscript{3} are highly concentrated and consistent across all three systems, suggesting minimal polarization stretching and stable transverse and longitudinal piezoelectric constants. In contrast, the distributions of $\Delta$\textit{P}\textsubscript{1} and $\Delta$\textit{P}\textsubscript{2} are more dispersed. However, after introducing defect dipoles, these distributions become more concentrated. This behavior suggests that defect dipoles hinder polarization rotation, thereby reducing the shear piezoelectric constant of KNbO\textsubscript{3}. The evolution of the standard deviation of polarization further confirms this change in distribution, indicating that the polarization vector is constrained by the defect dipoles. Thus, we conclude that defect dipoles impede polarization fluctuations in the non-polar direction, leading to a decrease in the shear piezoelectric constant.
\begin{figure}
    \centering
    \includegraphics[width=1\linewidth]{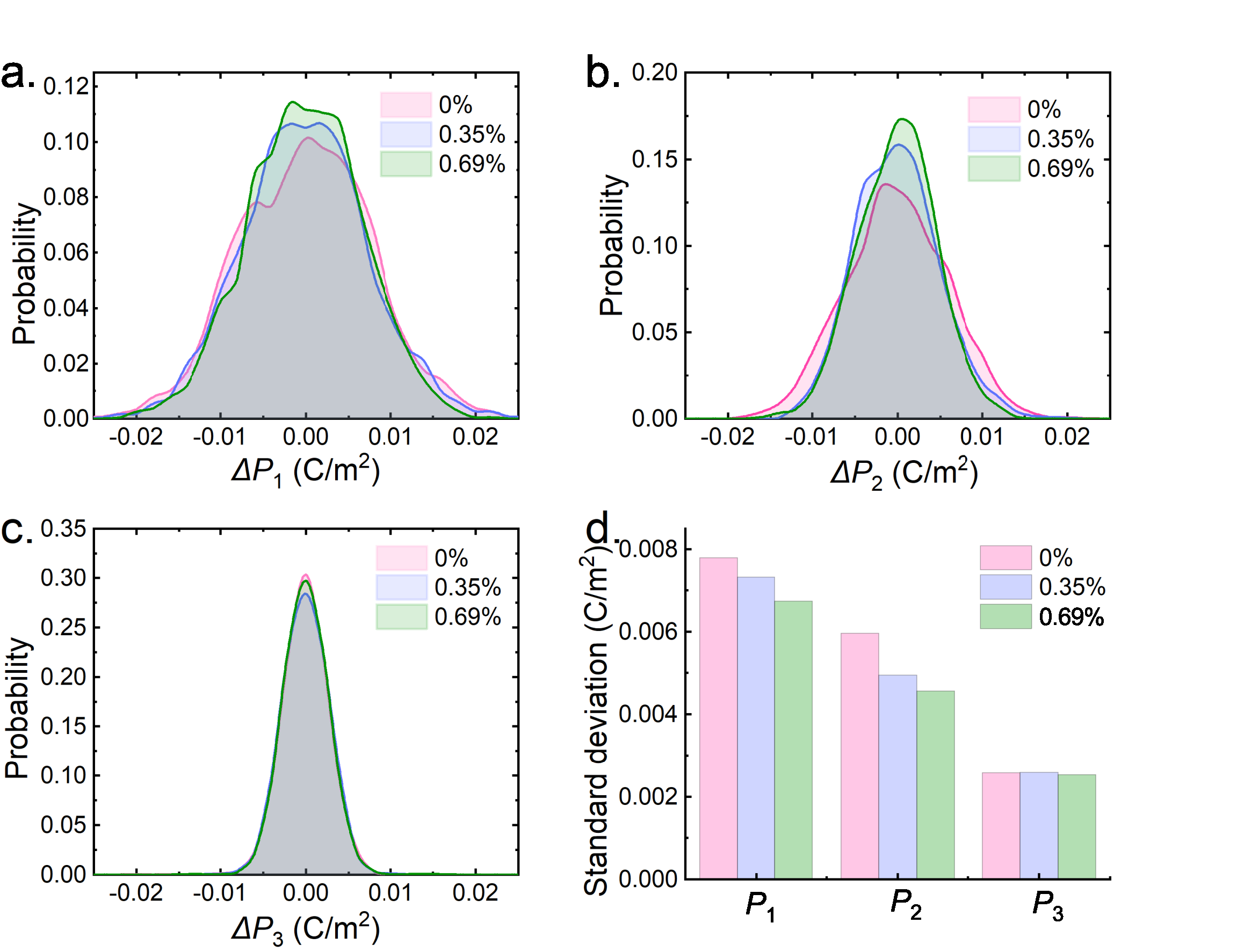}
    \caption{Distribution of polarization difference in (a) \textit{x}, (b) \textit{y}, and (c) \textit{z} direction; (d) Standard deviation of polarization fluctuation. }
    \label{Figure S3}
\end{figure}

\subsection{\label{sec:level1}\textbf{\textbf{\textbf{Orientational Average Method}
}}}
The calculations of the physical properties as a function of orientation can be done via coordinate transforms. We use a standard right-hand set with proper Euler angles to define the spatial orientation. Then the piezoelectric constant at arbitrary direction can be written as a linear combination of original piezoelectric matrix in tensor notation, 
\begin{equation}
d_{ijk}^{'}=a_{im}a_{jn}a_{kp}d_{ijk}^{0}\tag{S18}
\end{equation}
 where \textit{a}\textsubscript{\textit{ij}} is the transformation matrix elements and defined as the cosines of angles between the original axis $x_{i}^{0}$ in the Cartesian coordinate system \{$x_{1}^{0}$, $x_{2}^{0}$, $x_{3}^{0}$\} and the new axis $x_{j}^{'}$ in \{$x_{1}^{'}$, $x_{2}^{'}$, $x_{3}^{'}$\}, and the\textit{ i},\textit{ j}, \textit{k}, \textit{m}, \textit{n}, \textit{p} = \{\textit{x}, \textit{y}, \textit{z}\} in Cartesian directions, \textit{x} = 1, \textit{y} = 2, and \textit{z} = 3, full notation and the Einstein summation is used here. In our case, the elements $a_{ij}$ are written as
\begin{align}
a_{11}&=cos\varphi cos\psi-cos\theta sin\psi sin\varphi \nonumber\\
a_{22}&=cos\varphi sin\psi+cos\theta cos\psi sin\varphi \nonumber\\
a_{13}&=sin\varphi sin\theta \nonumber\\
a_{21}&=-sin\varphi cos\psi-cos\theta sin\psi cos\varphi \nonumber\\
a_{22}&=-sin\varphi sin\phi +cos\theta cos\psi cos\varphi \tag{S19} \\
a_{23}&=cos\varphi sin\theta \nonumber\\
a_{31}&=sin\theta sin\varphi \nonumber\\
a_{32}&=-sin\theta cos\varphi \nonumber\\
a_{33}&=cos\theta \nonumber
 \end{align}
where the  $\psi$, $\theta$, $\varphi$ are Euler angles, which describe the continuous and anticlockwise rotations in the ordinate system about the $x_{3}^0$ axis (by $\psi$),  $x_{1}^{0}$ axis (by $\theta$), and the final $x_{3}^{'}$ (by $\varphi$). We then noted that the orthorhombic KNbO\textsubscript{3} have a 12 possible equivalent polarization configuration. Supposing domain switching is completed during the poling process, the polarization of each grain should be induced in a nearest possible direction following the external electric field. Therefore, an orientational average in the region closest to the polarization axis can represent the piezoelectric characteristics of conventional KNbO\textsubscript{3} polycrystalline ceramics\cite{tan2019intrinsic}, as shown in Figure S4. Then the piezoelectric constant can be estimated by
\begin{equation}
    d_{ijk,c}=\frac{\iint_{R}sin\theta d\theta d\varphi\int_{0}^{2\pi}d_{ijk}^{'}d\psi}{\iint_{R}sin\theta d\theta d\varphi\int_{0}^{2\pi}d\psi} \tag{S20} 
\end{equation}
where the \textit{R} represents the region of double integral, which is shown in orange in Figure S4. 
\begin{figure}
    \centering
    \includegraphics[width=0.75\linewidth]{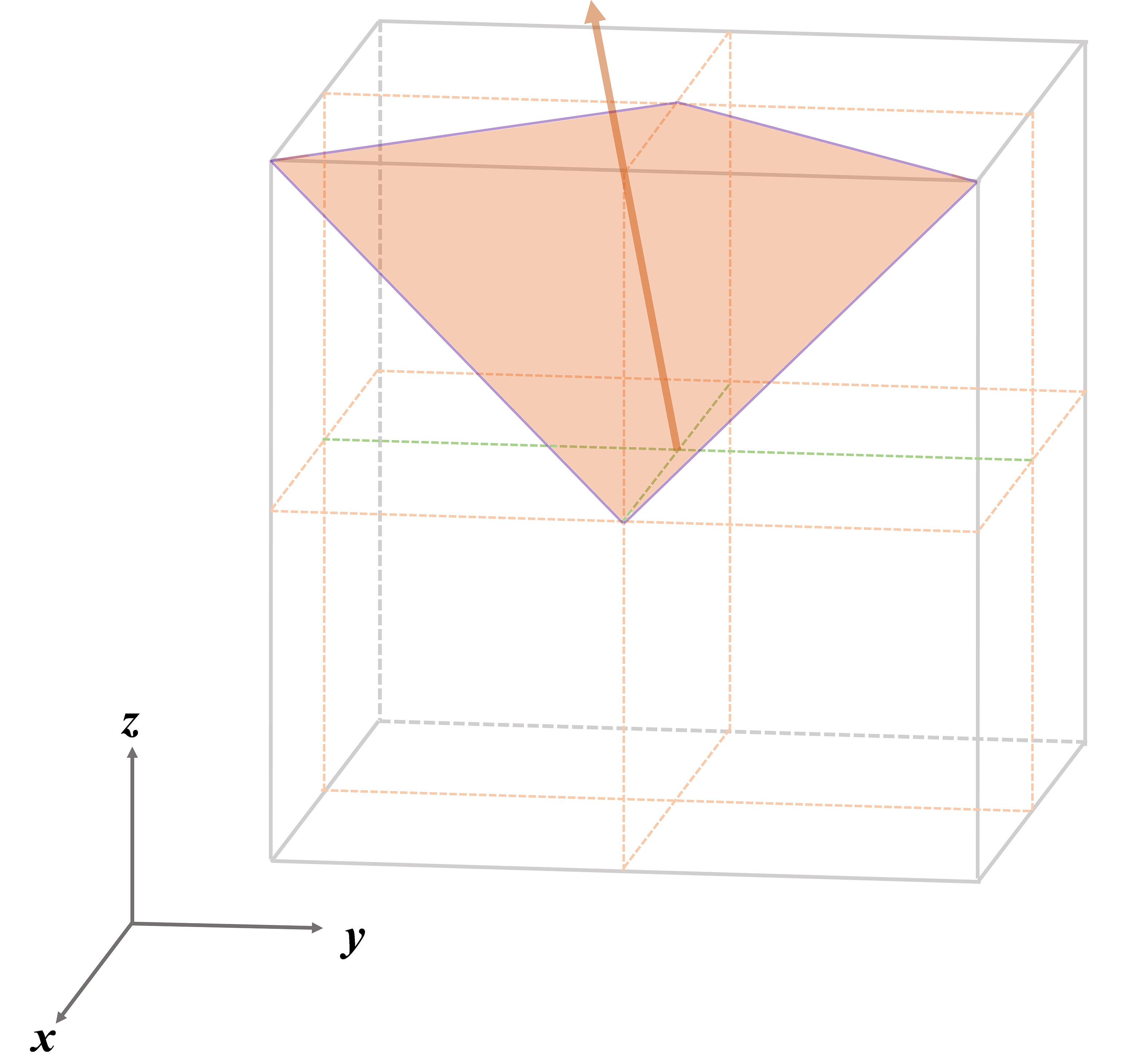}
    \caption{Schematic diagram of orientational average area in orthorhombic KNbO\textsubscript{3}, where the orange area is the possible region to induce [101] polarization. }
    \label{Figure S4}
\end{figure}
After calculations, we can obtain
\begin{align}
d_{31,c}&\approx 0.4128d_{31}+0.4272d_{32}+0.0722d_{33} \nonumber\\
&-0.0433d_{15}-0.0289d_{24} \nonumber\\
d_{33,c}&\approx 0.0866(d_{31}+d_{15})+0.0578(d_{32} +d_{24}) \nonumber \\
&+0.7678d_{33}  \tag{S21}\\
d_{15,c}&\approx -0.0433d_{31}-0.0289d_{32}+0.0722d_{33} \nonumber\\
&+0.4128d_{15}+0.4272d_{24} \nonumber
\end{align}
Equation (S21) can directly obtain the piezoelectric cosntants of polycrystalline ceramics from their properties of single crystal.

\bibliography{APS-Bending_SI}